\documentclass{llncs}

\usepackage{makeidx}  
\usepackage{amsmath}
\usepackage{mathrsfs}
\usepackage{amssymb}
\usepackage{algorithm}
\usepackage[noend]{algorithmic}
\usepackage{lipsum}
\usepackage{pstricks}
\usepackage{tikz}
\usepackage{subfigure}
\usepackage{wrapfig}
\usepackage{url}

\newtheorem{thm}{Theorem}

\begin{document}

\mainmatter              

\title{Fast approximate furthest neighbors with data-dependent hashing}

\titlerunning{Fast approximate furthest neighbors}  

\author{Ryan R. Curtin \and Andrew B. Gardner}

\authorrunning{Ryan R. Curtin and Andrew B. Gardner} 

\institute{Center for Advanced Machine Learning \\ Symantec Corporation \\
Atlanta, GA 30338, USA.\\
\email{ryan@ratml.org}, \email{Andrew\_Gardner@symantec.com}}

\maketitle
\begin{abstract}
We present a novel hashing strategy for approximate furthest neighbor search
that selects projection bases using the data distribution.  This strategy leads
to an algorithm, which we call {\tt DrusillaHash}, that is able to outperform
existing approximate furthest neighbor strategies.  Our strategy is motivated by
an empirical study of the behavior of the furthest neighbor search problem,
which lends intuition for where our algorithm is most useful.  We also present a
variant of the algorithm that gives an absolute approximation guarantee; to our
knowledge, this is the first such approximate furthest neighbor hashing approach
to give such a guarantee.  Performance studies indicate that {\tt DrusillaHash}
can achieve comparable levels of approximation to other algorithms while giving
up to an order of magnitude speedup.  An implementation is available in
the {\bf mlpack} machine learning library (found at
\url{http://www.mlpack.org}).
\end{abstract}

\section{Introduction}

We concern ourselves with the problem of {\it furthest neighbor search}, which
is the logical opposite of the well-known problem of nearest neighbor search.
Instead of finding the nearest neighbor of a query point, our goal is to find
the furthest neighbor.  This problem has applications in recommender systems,
where furthest neighbors can increase the diversity of recommendations
\cite{said2012increasing,said2013user}.  Furthest neighbor search is also a
component in some nonlinear dimensionality reduction algorithms
\cite{vasiloglou2008scalable}, complete linkage clustering
\cite{defays1977efficient,schloss2009introducing} and other clustering
applications \cite{veenman2002maximum}.  Thus, being able to quickly return
furthest neighbors is a significant practical concern for many applications.

However, it is in general not feasible to return exact furthest neighbors from
large sets of points.  Although this
is possible with Voronoi diagrams in 2 or 3 dimensions
\cite{cheong2003computing}, 
and with single-tree or dual-tree algorithms in higher dimensions
\cite{curtin2013tree}, these algorithms tend to have long running times in
practice.  Therefore, approximate algorithms are often considered acceptable in
most applications.

For approximate neighbor search algorithms, hashing strategies are a popular
option \cite{datar2004locality,indyk1998approximate,andoni2006near}.  Typically
hashing has been applied to the problem of nearest neighbor search, but recently
there has been interest in applying hashing techniques to furthest neighbor
search \cite{pagh2015approximate,indyk2003better}.  In general, these techniques
are based on random projections, where random unit vectors are chosen as
projection bases.  This allows probabilistic error guarantees, but the entirely
random approach does not use the structure of the dataset.

In this paper, we first consider the structure of the furthest neighbors
problem and then conclude that a data-dependent approach can be used to select
the projection bases for a hashing algorithm.  This allows us to develop:

\begin{itemize}
  \item {\tt DrusillaHash}, a hashing algorithm that uses data-dependent
projection bases and outperforms other approximate furthest neighbors approaches
in practice.

  \item A modified version of {\tt DrusillaHash} which satisfies rigorous
approximation guarantees, though it is not likely to be useful in practice.
\end{itemize}

Our empirical results in Section \ref{sec:experiments} show that the {\tt
DrusillaHash} algorithm demonstrably outperforms existing solutions for
approximate $k$-furthest-neighbor search.

\vspace*{-0.6em}
\section{Notation and formal problem description}
\vspace*{-0.5em}

The problem of furthest neighbor search is easily formalized.  Given a set of
{\it reference points} $S_r \in \mathcal{R}^{n \times d}$, a set of {\it
query points} $S_q \in \mathcal{R}^{m \times d}$, and a distance metric
$d(\cdot, \cdot)$, the problem is to find, for each query point $p_q \in S_q$,

\vspace*{-0.5em}
\begin{equation}
\operatorname{argmax}_{p_r \in S_r} d(p_q, p_r).
\end{equation}
\vspace*{-1.1em}

A trivial way to solve this algorithm is by brute-force: for each query
point, loop over all reference points and find the furthest one.  But this
algorithm takes $O(nm)$ time, and does not scale well to large $S_r$ or $S_q$.
In this paper, we will consider the $\epsilon$-approximate form of the furthest
neighbor search problem.

Given a set of {\it reference points} $S_r \in \mathcal{R}^{n \times d}$,
a set of {\it query points} $S_q \in \mathcal{R}^{m \times d}$, an
approximation parameter $\epsilon \ge 0$, and a distance metric $d(\cdot,
\cdot)$, the $\epsilon$-approximate furthest neighbor problem is to find a
furthest neighbor candidate $\hat{p}_{fn}$ for each query point $p_q \in S_q$
such that

\vspace*{-0.5em}
\begin{equation}
\frac{d(p_q, p_{fn})}{d(p_q, \hat{p}_{fn})} < 1 + \epsilon
\end{equation}
\vspace*{-0.5em}

\noindent where $p_{fn}$ is the true furthest neighbor of $p_q$ in $S_r$.  When
$\epsilon = 0$, this reduces to the exact furthest neighbor search problem.
This form of approximation is also known as relative-value approximation.

\vspace*{-0.5em}
\section{Related work}
\vspace*{-0.5em}

There have been a number of improvements over the naive brute-force search
algorithm suggested above.
Exact techniques based on Voronoi diagrams can solve the furthest neighbor
problem.  In 1981, Toussaint and Bhattacharya proposed building
a furthest-point Voronoi diagram to solve the furthest neighbors problem in $O(m
\log n)$ time \cite{toussaint1981geometric}.  But in high dimensions, Voronoi
diagrams are not useful because of their exponential memory dependence on the
dimension.

Another approach to exact furthest neighbor search uses space trees, as
described by Curtin et~al. \cite{curtin2013tree}.  A tree is built on the
reference points $S_r$, and nodes that cannot contain the furthest neighbor of a
given query point are pruned.  This is essentially equivalent to many algorithms
for nearest neighbor search, such as the algorithm for nearest neighbor search
with cover trees \cite{beygelzimer2006cover}, but with inequalities reversed
(i.e., we prune nearby nodes instead of faraway nodes).  It is also possible to
do this in a dual-tree setting, by also building a tree on the query points
$S_q$.  Dual-tree nearest neighbor search has been proven to scale linearly in
the size of the reference set under some conditions \cite{curtin2015plug};
however, no similar bound has been shown for dual-tree furthest neighbor search.
It would be reasonable to expect similar empirical scaling.  Unfortunately,
tree-based approaches tend to perform poorly in high dimensions, and the
construction time of the trees can cause the algorithm to be slower than
desirable in practice.

Further runtime acceleration can be achieved if approximation is allowed.  It is
easy to modify the single-tree and dual-tree algorithms to support this, in the
manner suggested by Curtin for nearest neighbor search \cite{curtin2015faster}.
Although this is shown to accelerate nearest neighbor search runtime by a
significant amount (depending on the allowed approximation), the setup time of
building the trees can still dominate.  A similar approach to this strategy is
the fair split tree, designed by Bespamyatnikh \cite{bespamyatnikh1996dynamic}.
But this approach suffers from the same issues.

The fastest known algorithms for approximate nearest neighbor search are hashing
algorithms.  Indyk \cite{indyk2003better} proposed a hashing algorithm based on
random projections that is able to solve a slightly different problem: this
algorithm is able to determine (approximately) whether or not there exists a
point in $S_r$ farther away than a given distance.  This can be reduced to the
approximate furthest neighbor problem we are interested in, but this is
complex to implement.

Pagh et al. \cite{pagh2015approximate} refine this approach in order to directly
solve the approximate furthest neighbor problem; this improves on the runtime of
Indyk's algorithm and is easy to implement.  This algorithm, called QDAFN
(`query-dependent approximate furthest neighbor'), has a guaranteed success
probability.  The algorithm is parameterized by the number of projections used
and the number of points stored for each projection; usually, this number is
relatively low.  But in extremely high-dimensional settings, the randomly-chosen
projections can fail to capture important outlying points.  This motivates us to
investigate the point distribution in order to choose projection bases.

\vspace*{-0.8em}
\section{Furthest neighbor point distribution}
\vspace*{-0.5em}

The furthest neighbor problem is significantly different from the nearest
neighbor problem, which has received significantly more attention
\cite{bentley1975multidimensional,arya1998optimal,gionis1999similarity,gray2000nbody,datar2004locality,curtin2013tree,curtin2015faster}.
This difference is perhaps somewhat counterintuitive, given that the furthest
neighbor problem is simply an $\operatorname{argmax}$ over all reference points
instead of an $\operatorname{argmin}$.  But this small change causes the problem
to have surprisingly different structure with respect to the results.

As a first observation of the differences between the two problems, consider
that for any set $S_r$, the furthest neighbor of every point can be made to be a
single point simply by adding a single point sufficiently far from every other
point in $S_r$.  There is no analog to this in the nearest neighbor search
problem.  Indeed, it is often true that for a furthest neighbor query with many
query points, the results may contain the same reference point.  This is easily
demonstrated.

Define the {\bf rank} of a reference point $p_r$ for some query point $p_q$ as
the position of $p_r$ in the ordered list of distances from $p_q$.  That is, if
the rank of $p_r$ for some query point $p_q$ is $k$, then $p_r$ is the
$k$-furthest neighbor from $p_q$.

\begin{figure}[b!]
  \centering
  \subfigure[{\tt cloud} dataset (10x2047).]{
    \includegraphics[width=0.47\textwidth]{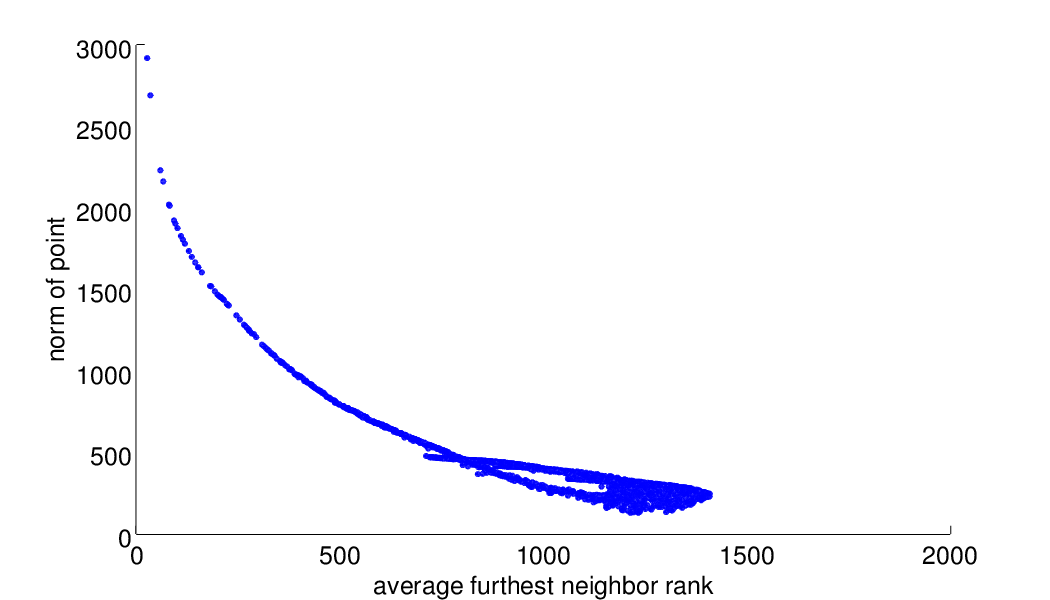}
    \label{fig:avgrank-cloud}
  }
  \subfigure[{\tt ozone} dataset (10x2047).]{
    \includegraphics[width=0.47\textwidth]{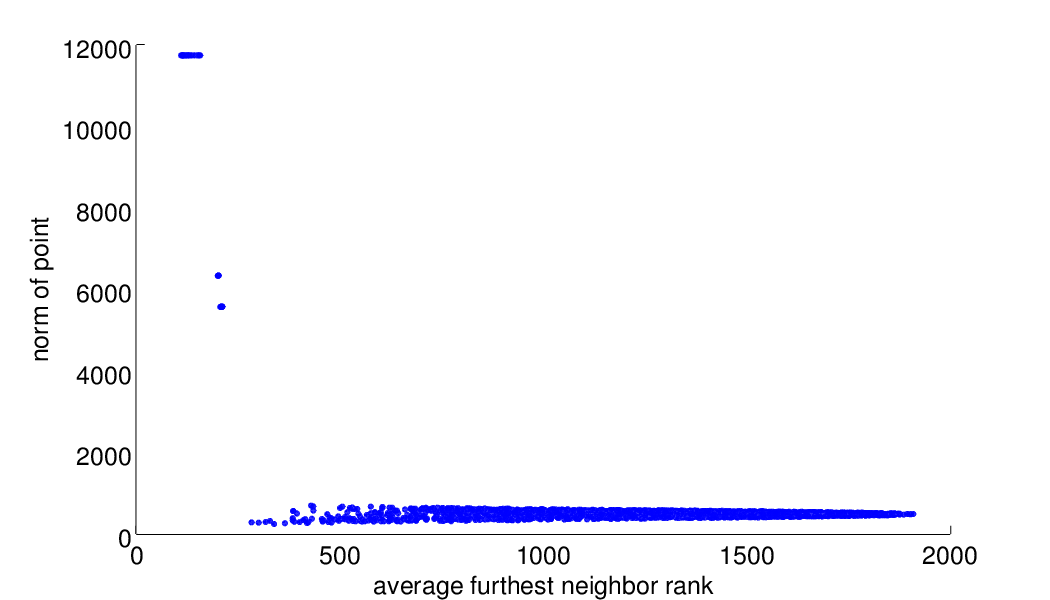}
    \label{fig:avgrank-ozone}
  }
  \subfigure[{\tt phy} dataset (78x150000).]{
    \includegraphics[width=0.47\textwidth]{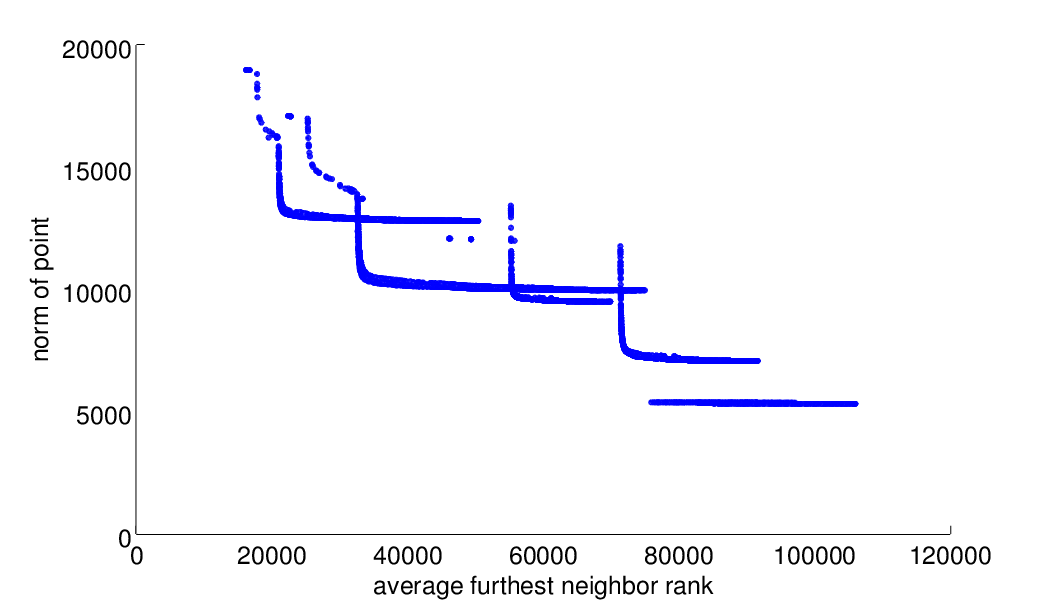}
    \label{fig:avgrank-phy}
  }
  \subfigure[{\tt covertype} dataset (55x581012).]{
    \includegraphics[width=0.47\textwidth]{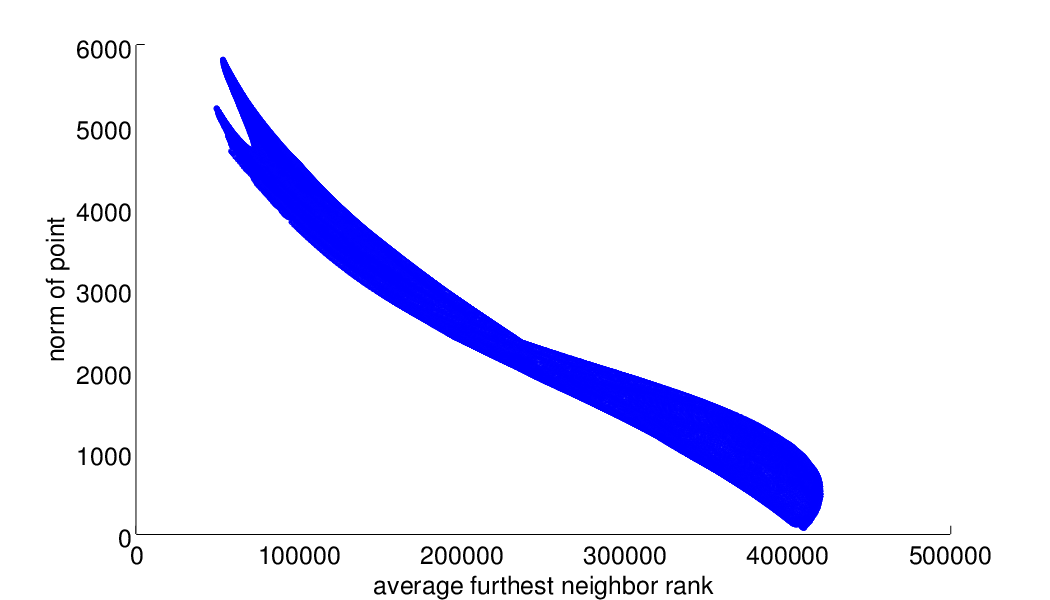}
    \label{fig:avgrank-covertype}
  }
  \caption{Average rank vs. norm for a handful of datasets.  Observe that a
large norm is correlated with a low rank.}
  \label{fig:avgrank}
  \vspace*{-1.0em}
\end{figure}

We can obtain insight into the behavior of furthest neighbor queries by
observing the average rank of points on some example datasets from the UCI
dataset repository \cite{ucimlrepository}.  Figure \ref{fig:avgrank} contains
scatterplots displaying the average rank of a reference point versus the
mean-centered norm of the reference point for the all-furthest-neighbors problem
(that is, each point in the reference set is used as a query point).

Figure \ref{fig:avgrank} shows that there is a clear and unmistakable
correlation between the norm of a point and its average rank for the
all-$k$-furthest-neighbor problem.  For the {\tt ozone} dataset, we can see that
there are only a few points with high norm, and all of these have much lower
average rank than the rest of the points.

These observations suggest that a reasonable approximate furthest neighbor
algorithm might be obtained simply by searching over the top few points in the
reference set with highest norm.  Unfortunately, an algorithm that simple will
fail in many cases in practice.  Still, an effective furthest neighbors
algorithm should take this structure into account: {\it high-norm points are
more important than low-norm points}.

\section{The algorithm: {\tt DrusillaHash}}
\vspace*{-0.5em}

\begin{figure}[t!]
  \centering
  \begin{tikzpicture}[scale=1.4]
  \coordinate (Origin) at (0, 0);
  \coordinate (XAxisMin) at (0, 0);
  \coordinate (XAxisMax) at (4, 0);
  \coordinate (YAxisMin) at (0, 0);
  \coordinate (YAxisMax) at (0, 3);
  \draw [thin, gray, -latex, ->] (XAxisMin) -- (XAxisMax);
  \draw [thin, gray, -latex, ->] (YAxisMin) -- (YAxisMax);

  \coordinate (v0) at (2.5, 1);
  \coordinate (p1) at (1.5, 2.5);
  \coordinate (p1proj) at (2.15517, 0.86207);

  \coordinate (a) at (1.9, 2.3);
  \coordinate (b) at (2.35517, 1.26207);

  \node [draw, circle, inner sep=1pt, fill] at (v0) { };
  \node [draw, circle, inner sep=1pt, fill] at (p1) { };
  \node [draw=none, fill=none, font=\scriptsize, right] at (v0) { $v_i$ };
  \node [draw=none, fill=none, font=\scriptsize, right] at (p1) { $p_j$ };

  \draw [black] (p1) -- (p1proj) node[draw=none, fill=none, font=\scriptsize,
midway, above, rotate=-68.2] { distortion };
  \draw [black] (Origin) -- (p1proj) node[draw=none, fill=none,
font=\scriptsize, midway, below, rotate=21.8] { offset };


  \draw [dashed, gray] (Origin) -- (v0);
\end{tikzpicture}
  \vspace*{-0.8em}
  \caption{Distortion and offset for $p_j$ with base vector $v_i$.}
  \vspace*{0.3em}
  \label{fig:offsetdistortion}
\end{figure}
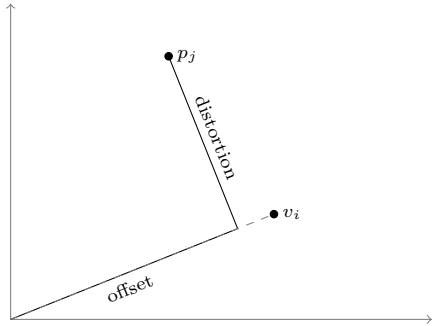

Our collective observations motivate a hashing algorithm for
approximate furthest neighbor search, which we introduce as {\tt
DrusillaHash} in Algorithm \ref{alg:drusillahash}.  The algorithm constructs
hash tables by repeatedly choosing points currently not in any hash table with
largest norm.\footnote{This is where the algorithm gets its name; the first
author's cat displays the same behavior when selecting a food bowl to eat from.}
After the hash tables are built, each query point is simply compared with all
points in each hash table in order to determine a good furthest neighbor
candidate.

\setlength{\textfloatsep}{0.6em}
\begin{algorithm}[t!]
\begin{algorithmic}[1]
  \STATE {\bf Input:} reference set $S_r$, query set $S_q$, number of neighbors
$k$, number of tables $l$, table size $m$
  \STATE {\bf Output:} array of furthest neighbors $N[]$
  \medskip

  \STATE \COMMENT{Pre-processing: mean-center data.}
  \STATE $m \gets \frac{1}{n} \sum_{p_r \in S_r} p_r$
  \STATE $S_r \gets S_r - m$; $S_q \gets S_q - m$ 
  \medskip
  \STATE \COMMENT{Pre-processing: build DrusillaHash tables.}
  \STATE $V \gets \{ \}$
  \STATE {\bf for all} $p_r \in S_r$ {\bf do} $n[p_r] \gets \| p_r \|$ \COMMENT{Initialize norms of points.}
  \FORALL{$i \in \{ 0, 1, \ldots, l \}$}
    \STATE $p_i \gets \operatorname{argmax}_{p_r \in S_r} n[p_r]$ \COMMENT{Take
next point with largest norm.} \label{alg:drusillahash:selectbasis}
    \STATE $v_i \gets p_i / \| p_i \|$
    \STATE $V \gets V \cup \{ v_i \}$
    \medskip

    \STATE \COMMENT{Calculate distortions and offsets.}
    \FORALL{$p_r \in S_r$ such that $n[p_r] \ne 0$}
      \STATE $O[p_r] \gets p_r^T v_i$
      \STATE $D[p_r] \gets \| p_r - O[p_r] v_i \|$
      \STATE $s[p_r] \gets | O[p_r] | - D[p_r]$ \label{alg:line:score}
    \ENDFOR
    \medskip

    \STATE \COMMENT{Collect points that are well-represented by $p_i$.}
    \STATE $R_i \gets$ points corresponding to largest $m$ elements of $s[\cdot]$
    \STATE {\bf for all} $p_r \in R_i$ {\bf do} $n[p_r] = 0$ \COMMENT{Mark point as used.}
    \FORALL{$p_r \in S_r$ such that $\operatorname{atan}(D[p_r] / O[p_r]) \ge
\pi / 8$}
      \STATE $n[p_r] = 0$ \COMMENT{Mark point as used.}
\label{alg:drusillahash:ignore}
    \ENDFOR
  \ENDFOR
  \medskip

  \STATE \COMMENT{Search for furthest neighbors.}
  \FORALL{$p_q \in S_q$}
    \FORALL{$R_i \in R$}
      \FORALL{$p_r \in R_i$}
        \IF{$d(p_q, p_r) > N_k[p_q]$}
          \STATE update results $N[p_q]$ for $p_q$ with $p_r$
        \ENDIF
      \ENDFOR
    \ENDFOR
  \ENDFOR
\end{algorithmic}
\caption{{\tt DrusillaHash}: fast approximate $k$-furthest neighbor search.}
\label{alg:drusillahash}
\end{algorithm}

{\tt DrusillaHash} depends on two parameters: $l$, the number of tables, and
$m$, the number of points taken for each table.  Empirically we observe that
values in the range of $l \in [2, 15]$ and $m \in [1, 5]$ produce acceptably
good approximations for most datasets, with approximation levels between
$\epsilon = 0.01$ and $\epsilon = 1.1$.

The primary intuition of the algorithm is that we want to collect points in the
hash tables $R_i$ that are likely to be furthest neighbors of any query point.
We know from our earlier experiments that points with high mean-centered
norms are likely to be good furthest neighbor candidates.  Thus, we start by
selecting the highest-norm mean-centered point $p_i$ as the primary point of the
table $R_i$, and collect $m$ points that are not too distorted by a projection
onto the unit vector $v_i$ which points in the direction of $p_i$.  Any points
that are not too distorted by this projection but not collected are ignored
for future tables (line \ref{alg:drusillahash:ignore}).

The words ``not too distorted'' deserve some elaboration: we wish to find
high-norm points that are well-represented by $p_i$, but we do not wish to find
high-norm points that are {\it not} well-represented by $p_i$.  Ideally, those
points will be selected as the primary point of another table $R_j$.  Therefore,
for each point $p_j$, we calculate the offset $O[p_j]$; this is the norm of the
projection of $p_j$ onto $v_i$.  Similarly, we calculate the distortion
$D[p_j]$.  Figure \ref{fig:offsetdistortion} displays a simple example of offset
and distortion.

Our goal is to balance two objectives in selecting points for $R_i$:

\vspace*{-0.4em}
\begin{itemize}
  \item Select high-norm points.
  \item Select points that are well-represented by $v_i$.
\end{itemize}
\vspace*{-0.4em}

\begin{figure}[t!]
  \centering
  \vspace*{-1.5em}
  \begin{tikzpicture}[scale=1.4]
  \coordinate (Origin) at (0, 0);
  \coordinate (XAxisMin) at (-3, 0);
  \coordinate (XAxisMax) at (3, 0);
  \coordinate (YAxisMin) at (0, -2);
  \coordinate (YAxisMax) at (0, 3);
  \draw [thin, gray, -latex, <->] (XAxisMin) -- (XAxisMax);
  \draw [thin, gray, -latex, <->] (YAxisMin) -- (YAxisMax);

  \coordinate (v0) at (2.5, 1);
  \coordinate (vn) at (-2.5, -1);
  \coordinate (p1) at (1.5, 2.5);
  \coordinate (p2) at (2.15, 1.2);
  \coordinate (p3) at (2.15, 0.7);
  \coordinate (p4) at (1, 2.4);
  \coordinate (p5) at (0.6, 0.7);
  \coordinate (p6) at (0.4, 1.1);
  \coordinate (p7) at (0.2, 0.2);
  \coordinate (p8) at (1.9, 0.2);
  \coordinate (p9) at (1.6, 1.0);
  \coordinate (p10) at (1.5, 0.7);
  \coordinate (p11) at (0.7, 1.6);
  \coordinate (p12) at (0.1, 0.9);
  \coordinate (p13) at (-0.6, 0.6);
  \coordinate (p14) at (0.6, -0.4);
  \coordinate (p15) at (-1.2, 0.1);
  \coordinate (p16) at (-1.6, -0.8);
  \coordinate (p17) at (-0.7, 1.1);
  \coordinate (p18) at (1.0, -0.7);
  \coordinate (p19) at (0.6, -1.3);
  \coordinate (p20) at (-0.3, -1.1);

  %

  \node [draw, circle, inner sep=1pt, fill] at (v0) { };
  \node [draw, circle, inner sep=1pt, fill=yellow] at (p1) { };
  \node [draw, circle, inner sep=1pt, fill=red] at (p2) { };
  \node [draw, circle, inner sep=1pt, fill=red] at (p3) { };
  \node [draw, circle, inner sep=1pt, fill=green] at (p4) { };
  \node [draw, circle, inner sep=1pt, fill=yellow] at (p5) { };
  \node [draw, circle, inner sep=1pt, fill=green] at (p6) { };
  \node [draw, circle, inner sep=1pt, fill=green] at (p7) { };
  \node [draw, circle, inner sep=1pt, fill=yellow] at (p8) { };
  \node [draw, circle, inner sep=1pt, fill=red] at (p9) { };
  \node [draw, circle, inner sep=1pt, fill=red] at (p10) { };
  \node [draw, circle, inner sep=1pt, fill=green] at (p11) { };
  \node [draw, circle, inner sep=1pt, fill=blue] at (p12) { };
  \node [draw, circle, inner sep=1pt, fill=blue] at (p13) { };
  \node [draw, circle, inner sep=1pt, fill=green] at (p14) { };
  \node [draw, circle, inner sep=1pt, fill=yellow] at (p15) { };
  \node [draw, circle, inner sep=1pt, fill=red] at (p16) { };
  \node [draw, circle, inner sep=1pt, fill=blue] at (p17) { };
  \node [draw, circle, inner sep=1pt, fill=blue] at (p18) { };
  \node [draw, circle, inner sep=1pt, fill=blue] at (p19) { };
  \node [draw, circle, inner sep=1pt, fill=green] at (p20) { };
  \node [draw=none, fill=none, font=\scriptsize, right] at (v0) { $v_i$ };
  \node [draw=none, fill=none, font=\scriptsize, right] at (p1) { $p_7$ };
  \node [draw=none, fill=none, font=\scriptsize, right] at (p2) { $p_1$ };
  \node [draw=none, fill=none, font=\scriptsize, right] at (p3) { $p_2$ };
  \node [draw=none, fill=none, font=\scriptsize, right] at (p4) { $p_{12}$ };
  \node [draw=none, fill=none, font=\scriptsize, right] at (p5) { $p_9$ };
  \node [draw=none, fill=none, font=\scriptsize, right] at (p6) { $p_{13}$ };
  \node [draw=none, fill=none, font=\scriptsize, right] at (p7) { $p_{10}$ };
  \node [draw=none, fill=none, font=\scriptsize, right] at (p8) { $p_6$ };
  \node [draw=none, fill=none, font=\scriptsize, right] at (p9) { $p_5$ };
  \node [draw=none, fill=none, font=\scriptsize, right] at (p10) { $p_4$ };
  \node [draw=none, fill=none, font=\scriptsize, right] at (p11) { $p_{11}$ };
  \node [draw=none, fill=none, font=\scriptsize, right] at (p12) { $p_{17}$ };
  \node [draw=none, fill=none, font=\scriptsize, right] at (p13) { $p_{18}$ };
  \node [draw=none, fill=none, font=\scriptsize, right] at (p14) { $p_{14}$ };
  \node [draw=none, fill=none, font=\scriptsize, right] at (p15) { $p_8$ };
  \node [draw=none, fill=none, font=\scriptsize, right] at (p16) { $p_3$ };
  \node [draw=none, fill=none, font=\scriptsize, right] at (p17) { $p_{19}$ };
  \node [draw=none, fill=none, font=\scriptsize, right] at (p18) { $p_{16}$ };
  \node [draw=none, fill=none, font=\scriptsize, right] at (p19) { $p_{20}$ };
  \node [draw=none, fill=none, font=\scriptsize, right] at (p20) { $p_{15}$ };


  \draw [dashed, gray] (Origin) -- (v0);
  \draw [dashed, gray] (Origin) -- (vn);
\end{tikzpicture}
  \vspace*{-0.8em}
  \caption{Example scores for a set of points; red: highest scores, blue: lowest
scores.}
  \label{fig:scores}
  \vspace*{-0.3em}
\end{figure}
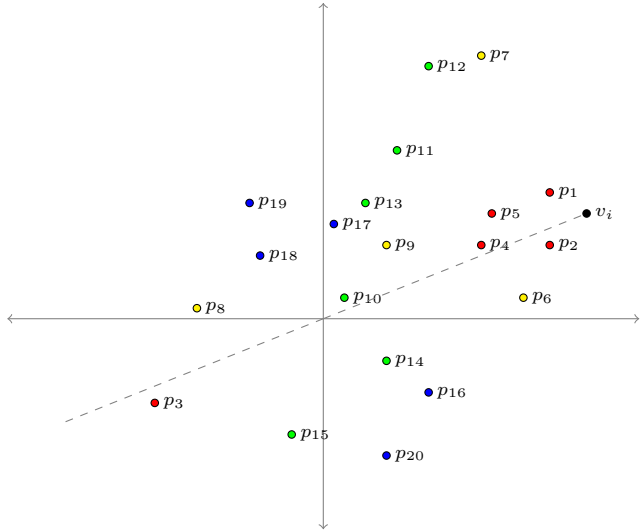

The solution we have used here is to construct a score $s[p_j]$ which is just
the distortion subtracted from the offset (see line \ref{alg:line:score}).
Figure \ref{fig:scores} displays an example $v_i$ with 20 points; each point is
indexed by its position in the ordered score set $s[\cdot]$.  In the context of
{\tt DrusillaHash}, if we took $m = 6$ (so, 6 points were selected for each
$v_i$), then $v_i$ and the five red points $p_1$ through $p_5$ would be selected
to make up the table $R_i$.  Then, $p_7$ would be chosen as $v_{i + 1}$ because
it is the point with largest norm that has not been selected to be in a hash
table (line \ref{alg:drusillahash:selectbasis}).

Once we have constructed the tables $R_i$, then our actual search is a simple
brute-force search over every point contained in each table $R_i$.  Because the
total number of points in $R$ is only $lm$, brute-force scan is sufficient.  In our experiments, attempting to prune points in $R$ involved too much
overhead.

{\tt DrusillaHash} has a similar structure to the
query-dependent approximate furthest neighbor algorithm of Pagh et~al.
\cite{pagh2015approximate} (``QDAFN''); except for three important differences: {\it (i)}
the vectors $v_i$ corresponding to each table are drawn using properties of the
reference set, {\it (ii)} there is no priority queue structure when scanning
the tables, and {\it (iii)} the projection bases chosen cannot be too similar.
Although {\tt DrusillaHash} can involve more setup time, our empirical
simulations show it is able to provide better results with fewer hash tables and
points in each hash table, resulting in better overall performance for a given
level of approximation.

\begin{table}[t!]
\centering
\begin{tabular}{|l|c|c|}
\hline
Algorithm & Setup time & Search time \\
\hline
{\tt DrusillaHash} & $O(ld |S_r| \log |S_r|)$ & $O(|S_q| dlm)$ \\
QDAFN \cite{pagh2015approximate} & $O(ld |S_r| \log |S_r|)$ & $O(|S_q| d(l \log
l + m \log l))$ \\
Indyk \cite{indyk2003better} & $O(ld |S_r| \log |S_r|)$ & $O(l |S_q|
(d + \log |S_r|) \log d \log\log d$\\
Brute-force & none & $O(|S_q| |S_r|)$ \\
\hline
\end{tabular}
\vspace*{0.3em}
\caption{Runtimes of approximate furthest neighbor algorithms.}
\label{tab:runtime}
\vspace*{-1.0em}
\end{table}

Table \ref{tab:runtime} gives a comparison of the runtimes of different
approximate furthest neighbor algorithms.  Note that {\tt DrusillaHash} and
QDAFN have the same asymptotic setup time for the same $l$ and $m$; but in
practice, the overhead of {\tt DrusillaHash} setup time is higher than QDAFN for
equivalent $l$ and $m$.  But again it must be noted that to provide the same
results accuracy, $l$ and $m$ may generally be set smaller with {\tt
DrusillaHash} than QDAFN.

\vspace*{-0.5em}
\section{Guaranteed approximation}
\vspace*{-0.2em}

Next, we wish to consider the problem of an absolute approximation guarantee: in
what situations can we ensure that the furthest neighbor returned is an
$\epsilon$-approximate furthest neighbor?

It turns out that this is possible with a modification of {\tt DrusillaHash},
given in Algorithm \ref{alg:guaranteedhash} as {\tt GuaranteedDrusillaHash}.
This algorithm, instead of taking a number of tables $l$, takes an acceptable
approximation level $\epsilon$.  The parameter $m$ does not affect the
theoretical results, and would only be interesting as an implementation detail.

\begin{algorithm}[t!]
\begin{algorithmic}[1]
  \STATE {\bf Input:} reference set $S_r$, query set $S_q$, number of neighbors
$k$, acceptable approximation level $\epsilon$, table size $m$
  \STATE {\bf Output:} array of furthest neighbors $N[]$
  \medskip

  \STATE \COMMENT{Pre-processing: mean-center data.}
  \STATE $m \gets \frac{1}{n} \sum_{p_r \in S_r} p_r$
  \STATE $S_r \gets S_r - m$; $S_q \gets S_q - m$ 
  \medskip

  \STATE \COMMENT{Pre-processing: build GuaranteedDrusillaHash tables.}
  \STATE $V \gets \{ \}$
  \FORALL{$p_r \in S_r$}
    \STATE $n[p_r] \gets \| p_r \|$ \COMMENT{Initialize norms of points.}
  \ENDFOR
  \STATE{$\delta \gets \frac{\epsilon}{15}$}
  \WHILE{$\max_{p_r \in S_r} n[p_r] > \delta \max_{p_r \in S_r} \| p_r
\|$}
    \STATE $p_i \gets \operatorname{argmax}_{p_r \in S_r} n[p_r]$ \COMMENT{Take
next point with largest norm.} \label{alg:guaranteedhash:select}
    \STATE $v_i \gets p_i / \| p_i \|$
    \STATE $V \gets V \cup \{ v_i \}$
    \medskip

    \STATE \COMMENT{Calculate distortions and offsets.}
    \FORALL{$p_r \in S_r$ such that $n[p_r] \ne 0$}
      \STATE $O[p_r] \gets p_r^T v_i$
      \STATE $D[p_r] \gets \| p_r - O[p_r] v_i \|$
      \STATE $s[p_r] \gets | O[p_r] | - D[p_r]$
    \ENDFOR
    \medskip

    \STATE \COMMENT{Collect points that are well-represented by $p_i$.}
    \STATE $R_i \gets\ $ points corresponding to largest $m$ elements of $s[\cdot]$
    \FORALL{$p_r \in R_i$}
      \STATE $n[p_r] = 0$ \COMMENT{Mark point as used.}
    \ENDFOR
  \ENDWHILE
  \medskip
  \STATE \COMMENT{Set shrug point (if we can).}
  \IF{there is any point such that $n[p_r] \ne 0$}
    \STATE $p_{sh} \gets $ some point such that $n[p_r] \ne 0$
  \ELSE
    \STATE $p_{sh} \gets \emptyset$
  \ENDIF

  \medskip
  \STATE \COMMENT{Search for furthest neighbors.}
  \FORALL{$p_q \in S_q$}
    \FORALL{$R_i \in R$}
      \FORALL{$p_r \in R_i$}
        \IF{$d(p_q, p_r) > N_k[p_q]$}
          \STATE update results $N[p_q]$ for $p_q$ with $p_r$
        \ENDIF
      \ENDFOR
    \ENDFOR

    \IF{$p_{sh} \ne \emptyset$ and $d(p_q, p_{sh}) > N_k[p_q]$}
      \STATE update results $N[p_q]$ for $p_q$ with $p_{sh}$
    \ENDIF
  \ENDFOR
\end{algorithmic}
\caption{{\tt GuaranteedDrusillaHash}: guaranteed approximate $k$-furthest
neighbor search.}
\label{alg:guaranteedhash}
\end{algorithm}

The algorithm is roughly the same as {\tt DrusillaHash}, except for that more
tables are added until all points with norm greater than $\delta
\max_{p_r \in S_r} \| p_r \|$ are contained in some hash table, and an extra
point called the {\it shrug point} is held.  The shrug point is set to be any
point within the small zero-centered ball of radius $\delta \max_{p_r \in S_r}
\| p_r \|$.  This is needed to catch situations where $p_q$ is close to every
point in $R_i$, and serves to provide a ``good enough'' result to satisfy the
approximation guarantee.

Because {\tt GuaranteedDrusillaHash} collects potentially huge numbers of hash
tables that may contain most of the points in $S_r$, the algorithm is primarily
of theoretical interest.  Although the algorithm will outperform
brute-force search as long as the hash tables do not contain nearly all of the
points in $S_r$, it is not likely to be practical for large $S_r$; thus, our
interest in {\tt GuaranteedDrusillaHash} is primarily theory-oriented.

With the algorithm introduced, we may present our theoretical result.  First, we
introduce a utility lemma.

\vspace*{-0.5em}
\begin{lemma}
Given a mean-centered set $S_r$ and a query point $p_q$ with true furthest
neighbor $p_{fn}$, if
$\| p_q \| < \frac{1}{3} \max_{p_r \in S_r} \| p_r \|$,
then
$\| p_{fn} \| > \frac{1}{3} \max_{p_r \in S_r} \| p_r \|$.
\label{lem:nottoosmall}
\end{lemma}
\vspace*{-1.2em}

\begin{proof}
This is a simple proof by contradiction: suppose $\| p_{fn} \| \le \frac{1}{3}
\max_{p_r \in S_r} \| p_r \|$.  Then, the maximum possible distance between
$p_q$ and $p_{fn}$ is bounded above as
$d(p_q, p_{fn}) < \frac{2}{3} \max_{p_r \in S_r} \| p_r \|$.
But the minimum possible distance between $p_q$ and the largest point in $S_r$
is bounded below as

\vspace*{-0.4em}
\begin{equation}
d(p_q, \underset{p_r \in S_r}{\operatorname{argmax}} \| p_r \|) \ge \max_{p_r \in S_r} \|
p_r \| - \frac{1}{3} \max_{p_r \in S_r} \| p_r \| = \frac{2}{3} \max_{p_r \in
S_r} \| p_r \|.
\end{equation}
\vspace*{-0.4em}

This means that the largest point in $S_r$ is a further neighbor than $p_{fn}$,
which is a contradiction. \qed
\end{proof}
\vspace*{-0.3em}

We may now prove the main result.

\begin{thm}
Given a set $S_r$ and an approximation parameter $\epsilon < 1$ and any table
size $m > 0$, {\tt GuaranteedDrusillaHash} will return, for each query point
$p_q$, a furthest neighbor $\hat{p}_{fn}$ such that

\vspace*{-0.8em}
\begin{equation}
\frac{d(p_q, p_{fn})}{d(p_q, \hat{p}_{fn})} < 1 + \epsilon
\end{equation}
\vspace*{-0.8em}

\noindent where $p_{fn}$ is the true furthest neighbor of $p_q$ in $S_r$.  That
is, $\hat{p}_{fn}$ is an $\epsilon$-approximate furthest neighbor of $p_q$.
\end{thm}
\vspace*{-0.6em}

\begin{proof}
We know from Lemma \ref{lem:nottoosmall} that if the norm of $p_q$ is less than
1/3 of the maximum norm of any point in $S_r$, then the true furthest neighbor
must have norm greater than 1/3 of the maximum norm of any point in $S_r$.
Since $\delta$ is always less than 1/3 in Algorithm \ref{alg:guaranteedhash}, we
know that any such point will be contained in some hash table $R_i$, and thus
the algorithm will return the exact furthest neighbor in this case.

The only other case to consider, then, is when the norm of the query point is
large: $\| p_q \| \ge \frac{1}{3} \max_{p_r \in S_r} \| p_r \|$.
But we already know due to the way the algorithm works, that if $\| p_{fn} \|
\ge \delta \max_{p_r \in S_r} \| p_r \|$,
then $p_{fn}$ will be contained in some hash table $R_i$ and the algorithm will
return $p_{fn}$, satisfying the approximation guarantee.

But what about when $\| p_{fn} \|$ is smaller?  We must consider the case where
$\| p_{fn} \| < \delta \max_{p_r \in S_r} \| p_r \|$.
Here we may place an upper bound on the distance between the query point and its
furthest neighbor:

\vspace*{-0.6em}
\begin{equation}
d(p_q, p_{fn}) \le \| p_q \| + \| p_{fn} \|
 < \| p_q \| + \delta \max_{p_r \in S_r} \| p_r \|.
\end{equation}
\vspace*{-0.9em}

We may also place a lower bound on the distance between the query point and its
returned furthest neighbor using the shrug point $p_{sh}$.  The distance between
$p_q$ and $p_{sh}$ is easily lower bounded:
$d(p_q, p_{sh}) \ge \| p_q \| - \delta \max_{p_r \in S_r} \| p_r \|$.
This is also a lower bound on $d(p_q, \hat{p}_{fn})$.  We may combine these
bounds:

\vspace*{-0.2em}
\begin{equation}
\frac{d(p_q, p_{fn})}{d(p_q, \hat{p}_{fn})} < \frac{\| p_q \| + \delta
\max_{p_r \in S_r} \| p_r \|}{\| p_q \| - \delta \max_{p_r \in S_r} \| S_r \|}.
\label{eqn:firsttogether}
\end{equation}
\vspace*{-0.4em}

Now, define the convenience quantity $\alpha$ as

\vspace*{-0.4em}
\begin{equation}
\alpha = \frac{\max_{p_r \in S_r} \| p_r \|}{\| p_q \|}.
\end{equation}
\vspace*{-0.4em}

Because of our assumptions on $p_q$, we know that $\alpha \le 3$.  This
also means that $\alpha^2 \le 3 \alpha$.  Similarly, we know that $\delta < 1$,
which means that $\delta^2 < \delta$.  Using these inequalities, we may further
simplify Equation \ref{eqn:firsttogether}.

\vspace*{-0.9em}
\begin{eqnarray}
\frac{d(p_q, p_{fn})}{d(p_q, \hat{p}_{fn})} &<& \frac{1 + \delta \alpha}{1 -
\delta \alpha} \\
 &=& \frac{1 + 2\delta\alpha + \delta^2 \alpha^2}{1 + \delta^2 \alpha^2} \\
 &<& 1 + 2\delta\alpha + \delta^2 \alpha^2 \\
 &<& 1 + 5\delta\alpha
\end{eqnarray}
\vspace*{-1.2em}

\noindent and since $\alpha \le 3$ and $\delta = \epsilon / 15$, then it is true
that

\vspace*{-0.4em}
\begin{equation}
\frac{d(p_q, p_{fn})}{d(p_q, \hat{p}_{fn})} < 1 + \epsilon
\end{equation}

\noindent and therefore the theorem holds. \qed
\end{proof}

Although {\tt GuaranteedDrusillaHash} does not guarantee better search time than
brute force under all conditions, it does in most conditions.  As one example,
consider a large dataset where
the norms of points in the centered dataset are uniformly distributed. Some of
these points will have norm less than $(\epsilon / 15) \max_{p_r \in S_r} \| p_r
\|$.  These points (except the shrug point $p_{sh}$) will not be considered by
the {\tt GuaranteedDrusillaHash} algorithm, and this means that the {\tt
GuaranteedDrusillaHash} algorithm will inspect fewer points at search time than
the brute-force algorithm.

Next, consider the extreme case, where there exists one outlier $p_o$ with
extremely large norm, such that the next largest point has norm smaller than
$(\epsilon / 15) \| p_o \|$.  Here, {\tt GuaranteedDrusillaHash} with $m = 1$
will only need to inspect two points: the extreme outlier, and the shrug point
$p_{sh}$.

On the other hand, there do exist cases where {\tt GuaranteedDrusillaHash} gives
no improvement over brute-force search, and every point must be inspected.  If
the dataset is such that all points have norm greater than $(\epsilon / 15)
\max_{p_r \in S_r} \| p_r \|$, then the tables $R_i$ will contain every single
point in the dataset.

These theoretical results show that it is possible to give a guaranteed
$\epsilon$-approximate furthest neighbor in less time than brute-force search,
if the distribution of norms of $S_r$ are not worst-case.  But due to the
algorithm's storage requirement, it is not likely to perform well in practice
and so we do not investigate its empirical performance.

\vspace*{-0.5em}
\section{Experiments}
\label{sec:experiments}
\vspace*{-0.5em}

Next, we investigate the empirical performance of the {\tt
DrusillaHash} algorithm, comparing with brute-force search, query-dependent
approximate furthest neighbor \cite{pagh2015approximate}, and dual-tree exact
furthest neighbor search as described by Curtin et~al. \cite{curtin2013tree} and
implemented in {\bf mlpack} \cite{curtin2013mlpack}.  Note that both brute-force
search and the dual-tree algorithm return exact furthest neighbors; {\tt
DrusillaHash} and QDAFN return approximations.

We test the algorithms on a variety of datasets from the UCI dataset repository
and {\tt randu}, which is uniformly randomly distributed points.  These datasets
and their properties are listed in Table \ref{tab:datasets}.  In addition,
hand-tuned parameters that produce $\epsilon = 0.05$-approximate furthest
neighbors (on average) are given for QDAFN and {\tt DrusillaHash}.

\begin{table}[b!]
\centering
\begin{tabular}{|l|c|c|c|c|c|c|c|}
\hline
 & & & \multicolumn{2}{|c|}{\ \ QDAFN params\ \ } & \multicolumn{2}{|c|}{\ \ {\tt
DrusillaHash} params\ \ } \\
{\bf Dataset} & $n$ & $d$ & \ \ \ \ \ \ $l$ \ \ \ \ \ \ & $m$ & \ \ \ \ \ \ \ $l$
\ \ \ \ \ \ \ & $m$ \\
\hline
cloud & 2048 & 10 & 30 & 60 & 2 & 1 \\
isolet & 7797 & 617 & 40 & 40 & 2 & 1 \\ 
corel & 37749 & 32 & 5 & 5 & 2 & 1 \\ 
randu & 100000 & 10 & 15 & 15 & 5 & 2 \\
miniboone & 130064 & 50 & 125 & 200 & 2 & 1 \\ 
phy & 150000 & 78 & 12 & 12 & 4 & 1 \\ 
covertype & 581012 & 55 & 15 & 20 & 6 & 2 \\ 
pokerhand\ \  & 1000000 & 10 & 15 & 50 & 50 & 8 \\ 
susy & 5000000 & 18 & 18 & 18 & 2 & 2 \\ 
higgs & \ 11000000 \ & 28 & 32 & 32 & 2 & 2 \\ 
\hline
\end{tabular}
\vspace*{0.3em}
\caption{Datasets and parameters.}
\label{tab:datasets}
\vspace*{-2.0em}
\end{table}

The first experiment is to compare runtimes across all four algorithms.  The
approximate algorithms are tuned to return $\epsilon = 0.05$-approximate
furthest neighbors (using the parameters from Table \ref{tab:datasets}).  Table
\ref{tab:runtimes} shows the average runtimes of each of the four algorithms on
each dataset across ten trials with the dataset randomly split into
30\% query set, 70\% reference set.  I/O times are not included; the runtime
only includes the time for the furthest neighbor search itself, including
preprocessing time (building hash tables or building trees).

\begin{table}[t!]
\centering
\begin{tabular}{|l|c|c|c|c|}
\hline
{\bf Dataset} & \ \ brute-force \ \ & \ \ dual-tree \ \ & \ \ QDAFN \ \ & \ \
{\tt DrusillaHash} \ \ \\
\hline
cloud & 0.0397s & 0.0404s & 0.010662s & {\bf 0.0013302s} \\
isolet & 6.7535s & 7.7057s & 0.16485s & {\bf 0.040634s} \\
corel & 10.292s & 1.030s & 0.021361s & {\bf 0.021122s} \\
randu & 42.392s & 28.004s & 0.31600s & {\bf 0.061855s} \\
miniboone & 187.26s & 4.1047s & 2.1648s & {\bf 0.10362s} \\
phy & 370.06s & 58.720s & 0.20293s & {\bf 0.18858s} \\
covertype & 4077.9s & 144.99s & 1.2439s & {\bf 0.20293s} \\
pokerhand \ \ & -- & 852.00s & 11.749s & {\bf 8.0353s} \\
susy & -- & 88.295s & 21.678s & {\bf 2.4467s} \\
higgs & -- & 425.05s & 56.094s & {\bf 12.694s} \\
\hline
\end{tabular}
\vspace*{0.3em}
\caption{Runtimes for $\epsilon = 0.05$-approximate furthest neighbor search.}
\label{tab:runtimes}
\vspace*{-1.0em}
\end{table}

The {\tt DrusillaHash} algorithm not only provides $\epsilon = 0.05$-approximate
furthest neighbors up to an order of magnitude faster than any other competing
algorithm, but it also needs to inspect fewer points to return an accurate
approximate furthest neighbor (with the exception of the {\tt pokerhand}
dataset).  In many cases, {\tt DrusillaHash} only needs to inspect fewer than 10
points to find good furthest neighbor approximations, whereas QDAFN must
inspect 50 or more.

Our datasets have two extreme examples: the {\tt miniboone} dataset, which is
high-dimensional but the data lies on a low-dimensional manifold, and the {\tt
randu} dataset, where points are uniformly distributed in the
10-dimensional unit ball.

For the {\tt miniboone} dataset, {\tt DrusillaHash} is able to easily recover
only four points that provide average 1.05-approximate furthest neighbors.  But
because QDAFN chooses random projection bases, it takes very many to have a
high probability of recovering good furthest neighbors.  In our experiments, we
were not able to achieve good approximation reliably until using as many as 125
projection bases.  This effect was also observed with the {\tt covertype}
dataset.

{\tt DrusillaHash} also outperforms other approaches on the {\tt randu} dataset,
despite there being no structure for {\tt DrusillaHash} to exploit.  But the
algorithm is still able to outperform others; this is because the algorithm
specifically ensures that projection bases are not too similar (see line
\ref{alg:drusillahash:ignore}).

\begin{figure}[b!]
\vspace*{0.6em}
\includegraphics[width=\textwidth]{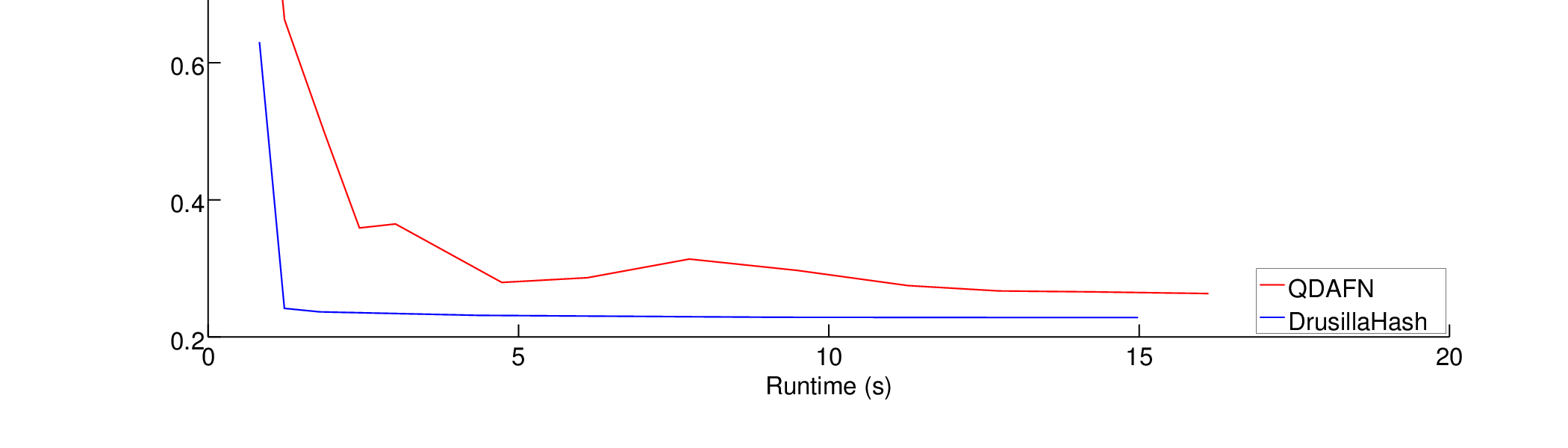}
\vspace*{-3.6em}
\caption{Maximum error for {\tt covertype} dataset as a function of runtime.}
\label{fig:covtype-scan}
\vspace*{-0.5em}
\end{figure}

Another important property of {\tt DrusillaHash} is that it gives a small
maximum error compared to QDAFN.  Figure \ref{fig:covtype-scan} shows the
maximum error of each approach as the number of points scanned increase on the
{\tt covertype} dataset.  For QDAFN, we have swept with $l = m$ from $l = 20$ to
$l = 250$, and for {\tt DrusillaHash}, we have set $m = l / 3$ and swept $l$
from $6$ to $60$.

Our experimental results have shown that {\tt DrusillaHash} gives excellent
approximation while only needing to scan few points.  Whereas QDAFN seems to
perform poorly in high-dimensional settings where the data lie on a
low-dimensional manifold (because projection bases are random), {\tt
DrusillaHash} effectively captures the low-dimensional structure with few
projection bases.

\vspace*{-0.4em}
\section{Conclusion}
\vspace*{-0.2em}

We have proposed an algorithm, {\tt DrusillaHash}, that builds hash tables for
approximate furthest neighbor search using the properties of the dataset to
choose the projection bases.  This algorithm design is motivated by our
empirical analysis of the structure of the approximate furthest neighbor search
problem, and the algorithm performs quite compellingly in practice.  It scales
better with dataset size than other techniques.

We have also proposed a variant, {\tt GuaranteedDrusillaHash}, which is able to
give an absolute approximation guarantee.  This is a benefit that no other
furthest neighbor hashing scheme is able to provide.  However, this variant is
not likely to be useful in practice.

Interesting future directions for this line of research may include combining a
random projection approach with the approach outlined here.  It would also be
possible to generalize our approach to arbitrary distance metrics, including
those where the points lie in an unrepresentable space.  This could be done
using techniques similar to some that have been used for max-kernel search
\cite{curtin2013fast,curtin2014dual}.

\vspace*{-0.5em}
\bibliographystyle{unsrt}
\bibliography{allkfn}

\end{document}